\relax
\documentclass[letterpaper]{article} 

\usepackage{caption}
\usepackage{xcolor,colortbl}
\usepackage{ulem}

\usepackage{aaai20}  
\usepackage{times}  
\usepackage{helvet} 
\usepackage{courier}  
\usepackage[hyphens]{url}  
\usepackage{graphicx} 
\usepackage{graphicx}  
\title{A Generalizable Method for Automated Quality Control of Functional Neuroimaging Datasets}
\author{Matthew Kollada, Qingzhu Gao, Monika S Mellem, Tathagata Banerjee, William J Martin \\ BlackThorn Therapeutics, San Francisco, CA 94103 \\ \textit{Email: matt.kollada@blackthornrx.com} }

\begin{document}
 \maketitle
 
\begin{abstract}

Over the last twenty five years, advances in the collection and analysis of functional magnetic resonance imaging (fMRI) data have enabled new insights into the brain basis of human health and disease. Individual behavioral variation can now be visualized at a neural level as patterns of connectivity among brain regions. As such, functional brain imaging is enhancing our understanding of clinical psychiatric disorders by revealing ties between regional and network abnormalities and psychiatric symptoms. Initial success in this arena has recently motivated collection of larger datasets which are needed to leverage fMRI to generate brain-based biomarkers to support the development of precision medicines. Despite numerous methodological advances and enhanced computational power, evaluating the quality of fMRI scans remains a critical step in the analytical framework. Before analysis can be performed, expert reviewers visually inspect individual raw scans and preprocessed derivatives to determine viability of the data. This Quality Control (QC) process is labor intensive, and the inability to adequately automate at large scale has proven to be a limiting factor in clinical neuroscience fMRI research. 

In this paper, we present a novel method for automating the QC of fMRI scans. We train machine learning classifiers using features derived from brain MR images to predict the ``quality'' of those images, which is based on the ground truth of an expert's opinion. Specifically, we emphasize the importance of these classifiers' ability to generalize their predictions across data collected in different studies. To address this, we propose a novel approach entitled ``FMRI preprocessing Log mining for Automated, Generalizable Quality Control'' (FLAG-QC), in which features derived from mining runtime logs are used to train the classifier. We show that classifiers trained on FLAG-QC features perform much better (AUC=0.79) than previously proposed feature sets (AUC=0.56) when testing their ability to generalize across studies. 

\end{abstract}
\section{Introduction}
\label{sec:introduction}
Brain imaging offers a window into the structure and function of the brain. Functional magnetic resonance imaging (fMRI) is a non-invasive method for measuring endogenous contrast in a blood oxygenation dependent (BOLD) signal as a proxy for brain activation.  Changes in the patterns of the BOLD signal can be used to track brain activity, during tasks or at rest, holding the promise of bringing objective measures of symptom severity to the field of clinical psychiatry which is a basis for precision medicine. For example, results from recent fMRI studies predicted individual symptom severity \cite{Mellem414037}, revealed therapeutic normalization of brain activity \cite{liu_ml_in_mdd}, and further subtyped patients within diagnostic categories \cite{Drysdale2016}. Recent empirical and simulation work has highlighted the importance of increasing the number of samples to maintain enough statistical power when analyzing fMRI data \cite{cremers}. Researchers have heeded these findings, as numerous large consortia have recently collected fMRI data at continuously growing orders of magnitude \cite{DiMartino2014}\cite{VANESSEN20122222}\cite{CASEY201843}. The UK Biobank exemplifies this new scale of brain imaging, having already reported collecting fMRI data, among other modalities, from 10,000 study participants with the stated intention of pushing that number to above 100,000 \cite{biobank}. However, with this new scale of data comes new challenges.

Raw fMRI images must undergo a complex set of computational transformations, often termed preprocessing, before being used in any statistical analysis. These raw and preprocessed images are commonly manually assessed for quality by expert reviewers in a process referred to as Quality Control (QC). These reviewers, often in multiple steps, visualize the preprocessed images, and inspect them for apparent errors that may erroneously bias future analysis \cite{QIMS25794}\cite{hitchhiker_fmri}. Many evaluation schemes for QC have been proposed \cite{biobank}\cite{mriqc}, however, one simple, clear strategy is determining whether the scan “passes” and is therefore usable, or “fails” and is discarded from further analysis. We use this scheme in our study.

The labor-intensive and time-consuming nature of QC serves as a clear bottleneck to the analysis of fMRI images at scale. QC of an fMRI dataset with hundreds of scans can take weeks to months of manual assessment from a single expert reviewer before analysis can begin. As discussed above, many recent fMRI studies have collected data at or even above that scale \cite{biobank}\cite{VANESSEN20122222}\cite{DiMartino2014}\cite{CASEY201843}, providing compelling motivation for the field to develop a scalable QC framework to reduce the burden on individual researchers and standardize quality control of fMRI data. 

One tangible, potential solution is to train machine learning classifiers to predict image quality based on expert reviewer ground-truth QC labels. This could massively decrease manual reviewer QC time and overcome the current bottleneck. Researchers have made certain progress in creating automatic QC classifiers for structural Magnetic Resonance Imaging (sMRI), which examines the anatomical structure of the brain and must also undergo a QC process. In these paradigms, feature sets derived from the raw and preprocessed sMRI images have been used to train machine learning classifiers to predict manual QC labels. Here, Esteban et. al \shortcite{mriqc} made important progress, testing a predictive framework for automatic sMRI QC, and importantly, evaluating the model's ability to generalize across datasets.

Recent literature, however, demonstrates that fMRI can either successfully augment or individually outperform sMRI in many crucial neuropsychiatric applications \cite{Mellem414037} \cite{Liu453951} \cite{ml_in_depression}. Nonetheless, little effort has been made to apply the QC framework from sMRI to fMRI, providing clear incentive to develop automatic QC classification on fMRI data. 

In this study we present our work in building automatic QC classifiers for fMRI data. Using two large, open-source fMRI datasets we build these classifiers and evaluate a range of feature sets, including one novel to this work entitled, “FMRI preprocessing Log mining for Automated, Generalizable Quality Control,” or FLAG-QC. Specifically, we evaluate the ability of these classifiers to generalize across fMRI data collected within different studies.

Our results demonstrate that we are able to achieve this generalization using only the novel FLAG-QC feature set proposed within this work.

\begin{figure}[t]
    \centering
    \includegraphics[width=\linewidth]{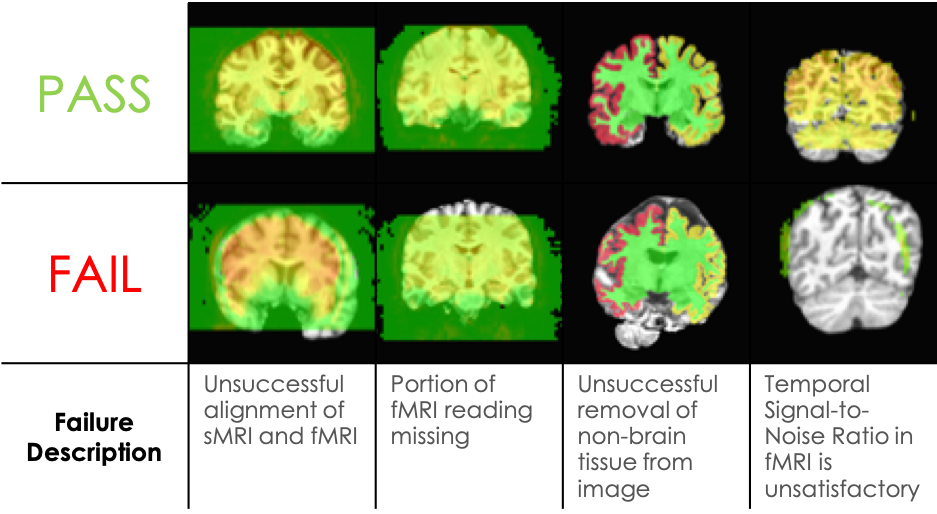}
    \caption{Examples of 2D Slices of preprocessed fMRI Images labelled as both passing and failing by expert QC reviewers}
    \label{fig:qc_example}
\end{figure}

\section{Background}

Automation of QC for brain imaging scans has received attention in structural MRI. Some of the first approaches attempted to quantify measures of image quality by creating indices, often termed ``Image Quality Metrics'' (IQMs), through which the quality of a raw sMRI scan can be judged in comparison to other scans. Woodard and Carley-Spencer \shortcite{Woodard2006} explored the adequacy of IQMs derived from Natural Scene Statistics and those defined by the JPEG Consortium. Their work identified the ability of IQMs to discriminate between raw structural MR images and those that had been artificially distorted. Similarly, Mortamet et. al \shortcite{mortamet} proposed a pair of sMRI quality indices based on the analysis of the background air (non-brain) volumetric pixels (voxels) of a brain sMRI scan, which proved to be useful in identifying poor quality scans in Alzheimer patients. The above-mentioned studies facilitated the understanding of certain attributes or features that are able to contribute to the discrimination between high and low quality scans, but stopped short of testing of a predictive framework evaluated on data held-out during training.

Pizzaro et. al \shortcite{pizzaro} built on this approach by introducing their own set of features, engineered to target specific artifacts found in brain sMRI scans. These features were used to train a Support Vector Machine (SVM) classifier to predict manual expert QC labels on 1457 3D-MRI volumes within cross validation experiments. Similarly, Alfaro-Almagro et. al \shortcite{biobank}, in working with the first set of imaging data to come out of the UK Biobank study, created automatic QC classifiers with ~5800 structural scans. They developed a set of 190 features to train an ensemble machine learning classifier to predict manual QC labels. Together, these studies expanded the space of features explored in automatic MRI QC and tested predictive modeling frameworks. However, neither evaluated the generalizability of their classifiers across datasets.

Esteban et. al \shortcite{mriqc} illuminated a limitation in across-dataset generalizability, testing whether classifiers could predict QC labels when trained on a dataset from one study and evaluated on data from an entirely different one. Their aim was to successfully train an sMRI QC classifier with scans from the Autism Brain Imaging Data Exchange \cite{DiMartino2014} (ABIDE) (N=1001) dataset to be evaluated on scans from the UCLA Consortium for Neuropsychiatric Phenomics LA5c \cite{CNP} (CNP) (N=265) dataset. The study revealed that their classifier performed quite well in cross-validation prediction within ABIDE. However, the model's predictive power dropped notably when trained on ABIDE and tested across datasets on CNP. This work brought to light the challenge of building robust automatic QC prediction of any MRI data.

As a whole, the study of automatic MRI QC has made progress, but still faces notable limitations. The generalizability of QC classifiers to new datasets must improve if practical applications are to be built on top of these methods. Further, as discussed in the \textbf{Introduction}, these methods have not been tested within fMRI data even though it has proved itself as a powerful research modality. To address these limitations we evaluate the existing automatic QC frameworks within fMRI, in addition to exploring new feature spaces and modeling paradigms.

\section{Data Sources}
To demonstrate the effectiveness of our methods, we use fMRI scans obtained from two separate studies: 1) Establishing Moderators and Biosignatures of Antidepressant Response for Clinical Care for Depression \cite{embarc} (EMBARC), 2) UCLA Consortium for Neuropsychiatric Phenomics LA5c \cite{CNP} (CNP). The CNP study was also used in previously mentioned work of Esteban et. al \shortcite{mriqc}.

\subsection{EMBARC}

The EMBARC dataset was collected to examine a range of biomarkers in patients with depression to understand how they might be able to inform clinical treatment decisions. The study enrolled 336 patients aged 18-65, collecting demographic, behavioral, imaging, and wet biomarker measures for multiple visits over a period of 14 weeks. Data were acquired from the National Data Archive (NDA) repository on June 19, 2018 with a license obtained by Blackthorn Therapeutics.

Our study only analyzes data from sMRI and fMRI scans collected during patients' first and second visit to the study site. Specifically, we use T1-weighted structural MRI scans and T2*-weighted blood-oxygenation-level-dependent (BOLD) resting-state functional MRI scans that were labelled as “run 1”. In total we analyze 324 structural-functional MRI scan pairs from the first site visit and 288 pairs from the second, producing a total of 612 scan pairs.

\subsection{CNP}

The CNP dataset was collected to facilitate discovery of the genetic and environmental bases of variation in psychological and neural system phenotypes, to elucidate the mechanisms that link the human genome to complex psychological syndromes, and to foster breakthroughs in the development of novel treatments for neuropsychiatric disorders. The study enrolled a total of 272 participants aged 21-50. Within the participant group, there were 138 healthy individuals, 58 diagnosed with schizophrenia, 49 diagnosed with bipolar disorder, and 45 diagnosed with ADHD. All data were collected in a single visit per participant and included demographic, behavioral, and imaging measures. We used version 1.0.5 of the study data downloaded from openneuro.org on January 4, 2018 with a license obtained by Blackthorn Therapeutics.

Similar to EMBARC, we use data from participants that have both T1-weighted sMRI and T2*-weighted BOLD resting-state fMRI scans that were labelled “run 1”. This amounts to 251 structural-functional MRI scan pairs.

\section{Methods}

We define the problem addressed in this study within a classification framework. Using features derived from brain fMRI scans, we train machine learning classifiers that predict QC outcome labels provided by expert reviewers. In this section, we briefly discuss each of the steps we take in our study. 

\begin{figure}[t]
    \centering
    \includegraphics[width=\linewidth]{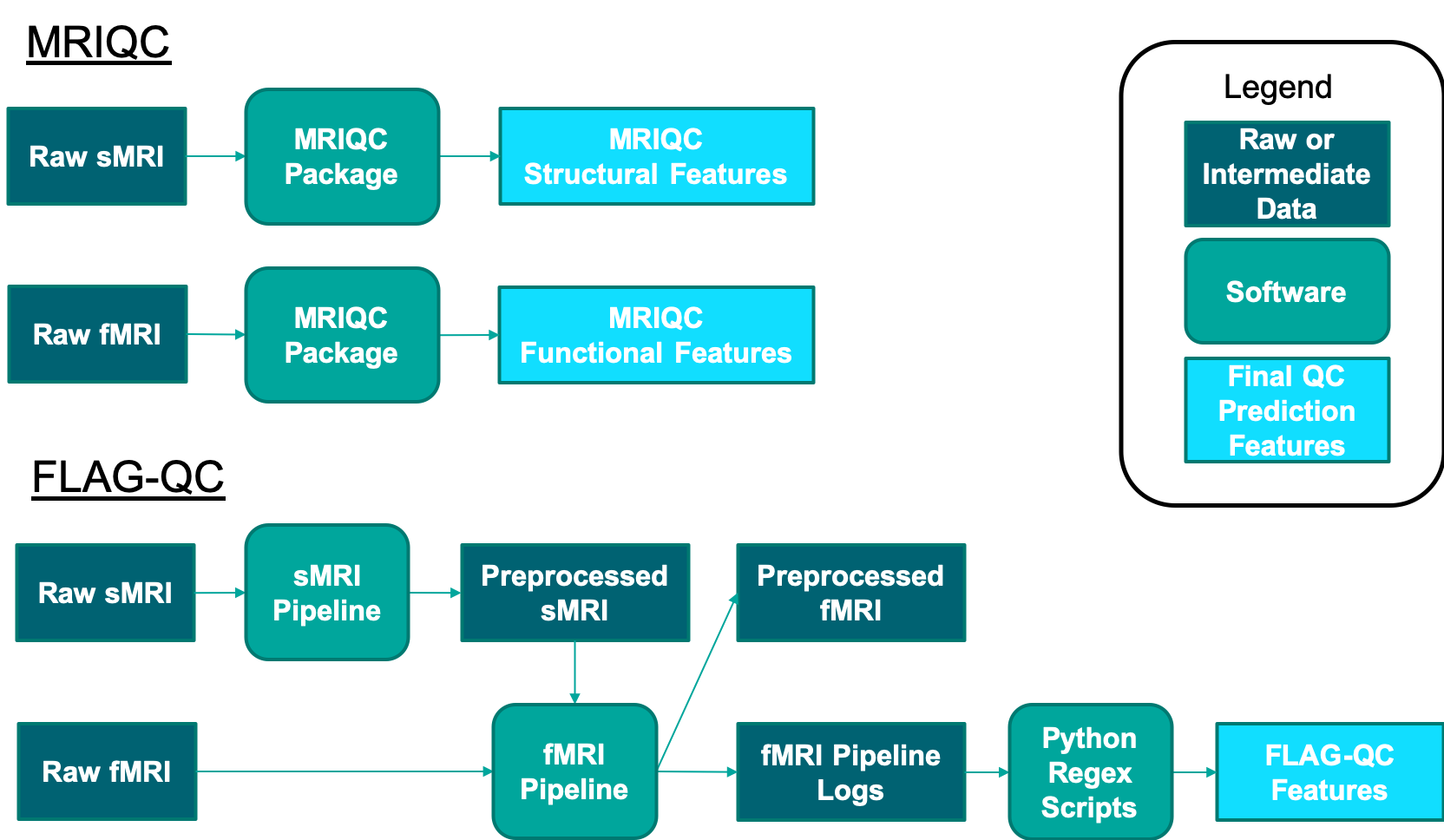}
    \caption{Flow charts detailing the processes for creating the individual feature sets used in automatic QC prediction}
    \label{fig:feature_generation}
\end{figure}

\subsection{MRI Preprocessing Pipeline}

In the study of brain MRI, preprocessing refers to a set of transformations that a raw image must undergo to prepare it for data analysis. All MRI data used in our analysis was preprocessed by a pipeline with two distinct modules, one for preprocessing T1-weighted sMRI scans and one for T2*-weighted resting-state fMRI scans. The sMRI module uses the FreeSurfer Software Suite \cite{reuter_2013}\cite{FISCHL2002341}, specifically the \textit{recon-all} command, to segment the structural image into various structures of the brain (parcellated gray and white matter regions of the cortex and subcortical regions). The fMRI module then uses the Analysis of Functional Neuroimages (AFNI) Software Suite \cite{COX1996162} to preprocess the raw resting-state fMRI image and the sMRI image further. This pipeline includes steps to remove noise in collected images, standardize scans to a template space for across scan comparison, segment the brain image into different tissue types and regions of interest, and extract numerical features from the preprocessed brain image for further analysis (see \cite{Mellem414037} for the full details of our sMRI and fMRI preprocessing steps).

\subsection{Feature Sets}

The features that we use to train QC classifiers come from two distinct pipelines: 1) FLAG-QC Features, a feature set novel to this study, and 2) MRIQC Features, those generated by the \textit{MRIQC} software suite \cite{mriqc}. A high-level block diagram showing the process for creating each set of features is shown in Figure \ref{fig:feature_generation}.

\begin{figure}[ht]
    \centering
    \fbox{
        \includegraphics[width=.95\linewidth]{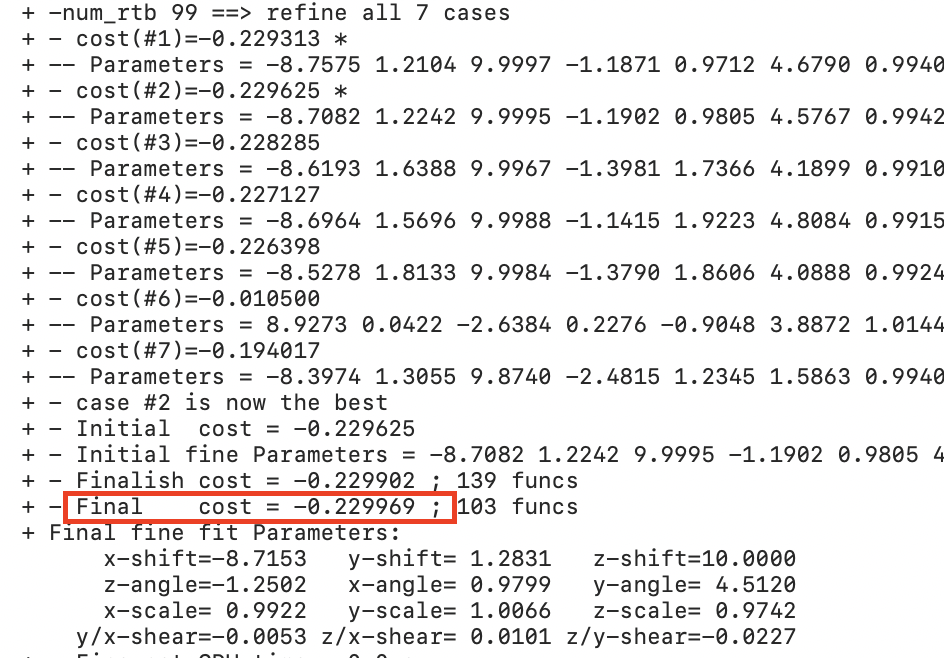}
    }
    \caption{A portion of an individual scan's fMRI preprocessing log text file. The highlighted region displays a FLAG-QC feature mined from this log file for each scan..}
    \label{fig:log_snapshot}
\end{figure}

\subsubsection{FLAG-QC Features}

The FLAG-QC features are derived from AFNI software commands run within our fMRI preprocessing pipeline. These commands are responsible for transforming the fMRI image into the final outputs that undergo manual QC, making information about their outcomes useful for automated QC decisions. While an AFNI command is executing, it emits runtime logs. Our preprocessing pipeline copies those logs and saves them to text files. These logs contain a large assortment of information, some of it pertaining to results of final or intermediate steps of a given command. We use the Python Regular Expression library to parse the text files and extract potentially informative features. An example of a portion of an fMRI pipeline log file is shown in Figure \ref{fig:log_snapshot}. 

The collected FLAG-QC Features can be divided into four subgroups; Step Runtimes, Voxel Counts, Brain Coordinates, and Other Metrics. 1) Step Runtime features quantify how long a given step, or set of steps, in the pipeline took to run. 2) Voxel Count features measure the size of the output of a given step in the pipeline in terms of ``voxels'', or volumetric 3D pixels. 3) Brain Coordinate features simply refer to the X, Y, and Z coordinates of the bounding box of the brain image. 4) Other Metrics are miscellaneous values that quantify the outcome of a certain step of the preprocessing pipeline. An example of one of these Other Metrics is the cost function value associated with the step of the pipeline that aligns the structural and functional scans. In total, there are 38 FLAG-QC features.

The full explanation of each individual feature falls outside the scope of this paper, but the authors welcome contact to discuss the features further.

\subsubsection{MRIQC Features}

\textit{MRIQC} is a software developed by the Poldrack Lab at Stanford University \cite{mriqc}. One of its features is the ability to generate measures of image quality from raw MRI images. These Image Quality Metrics (IQMs) are used in Esteban et. al \shortcite{mriqc} to predict manual QC labels on sMRI scans. The metrics are designated as ``no-reference'', or having no ground-truth correct value. Instead, the metrics generated from one image can be judged in relation to a distribution of these measures over other sets of images. \textit{MRIQC} generates IQMs from both structural and functional raw images. 

The structural IQMs are divided into four categories: measures based on noise level, measures based on information theory, measures targeting specific artifacts, and measures not covered specifically by the other three. 

The functional IQMs are broken down into three categories: measures for spatial structure, measures for temporal structure, and measures for artifacts and others. 

In total there are 112 features generated by \textit{MRIQC}, 68 structural features and 44 functional features. A full list of the features generated by \textit{MRIQC} can be found at mriqc.readthedocs.io. The software can be run as either a Python library or Docker container. We used the Docker version to generate IQMs on EMBARC and CNP.

\subsection{Feature Selection}

In this work we use a two-phase feature-selection approach:
\begin{enumerate}
\item In the first phase, we apply a model-independent approach. Specifically we use Hilbert-Schmidt Independence Criterion Lasso (HSIC Lasso) based feature selection \cite{Yamada_2014}. HSIC Lasso utilizes a feature-wise kernelized Lasso for capturing non-linear input-output dependency. A globally optimal solution can be efficiently calculated making this approach computationally inexpensive.
\item In the second phase, we apply a model-dependent Forward Selection approach \cite{sfs}.
\end{enumerate}
We choose this two-phase approach because it offers a good balance of classifier performance, fast computation and generalization. The actual number of features selected depends on cross-validation performance.

\subsection{Expert Manual QC Labels}

Our classifiers attempt to predict QC outcome labels that come from a trained reviewer who has years of experience working with MRI data. It is common practice in the field that expert QC reviewers examine raw MRI scans and preprocessed images to determine if the quality is sufficient for further analysis \cite{QIMS25794}\cite{hitchhiker_fmri}. For volumetric data, the 3D preprocessed MRI images are spatially sampled as 2D images for easier assessment by the reviewer. Examples of 2D images that both ``pass'' and ``fail'' QC are shown in Figure \ref{fig:qc_example} with common failure points, such as mis-alignment of structural and functional MRI scans or unsuccessful automatic removal of non-brain tissue, highlighted. In our study, following assessment of image quality of raw data and across the multiple preprocessing steps, the reviewer makes a binary ``pass'' or ``fail'' decision for each subject's fMRI scan.  Thus, an fMRI scan is tagged as useable (pass) or not (fail), and these labels serve as the ground-truth decisions on which our classifier is trained. 

\begin{figure*}[hbt!]
    \centering
    \begin{minipage}{\linewidth}
    \includegraphics[width=.48\linewidth]{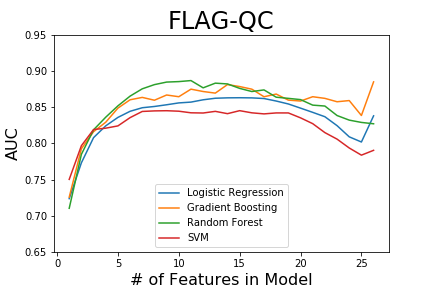}
    \includegraphics[width=.48\linewidth]{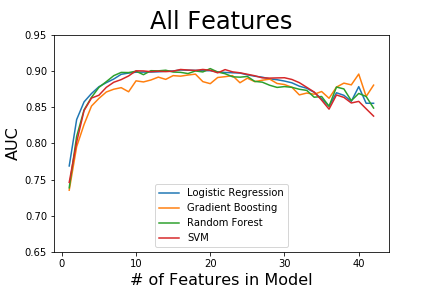}
   \\
    \includegraphics[width=.48\linewidth]{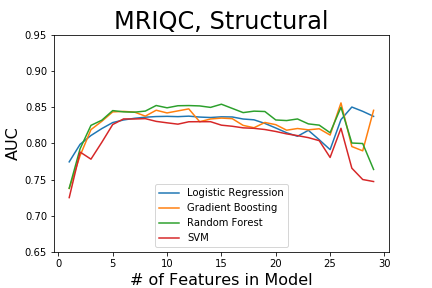}
     \includegraphics[width=.48\linewidth]{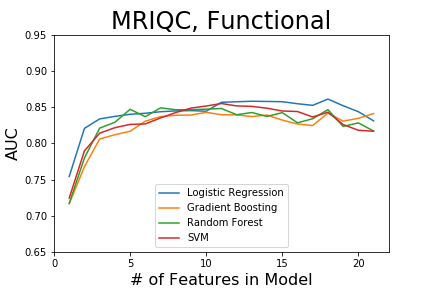} 
    \end{minipage}
    \captionsetup{justification=centering}
    \caption{Forward Feature Selection Classification Results on Set of Features Chosen by HSIC Lasso - EMBARC }
    \label{fig:emb_sfs_results}
\end{figure*}

For the data presented here, the ground-truth QC outcome labels were generated by a single expert reviewer who performed QC on 612 scans from EMBARC and 251 scans in CNP. Our expert reviewer labelled 73.9\% (452/612) of the scans from EMBARC as passing QC and 58.6\% (147/251) as passing for CNP. 

Inter-rater reliability has been noted as a source of label noise in other automatic MRI QC efforts \cite{mriqc}. To assess this possible noise in our data, a second expert reviewer independently performed QC on the 251 scans from CNP. The reviewers' pass/fail QC decisions agreed on 87.6\% (220/251) of the scans. That there is some disagreement is not surprising, given that the current standard of manual fMRI QC has certain subjective aspects, but the overall high congruence of the two raters predictions gives us confidence that reasonable predictions can be made on these labels.

Going forward, we will use only the labels of the first reviewer mentioned, who performed QC on both EMBARC and CNP, to train and test our classifiers.

\section{Results}

In this section, we demonstrate that our classifiers can accurately predict manual QC labels on fMRI scans within one data source using any of the feature sets mentioned above, but that only the novel FLAG-QC feature set proposed in this study successfully generalizes to data of another independent study. To address these two scenarios, we structure our results in two folds. Data collected from the same study will be referred to as ``within dataset'' samples, while data collected from a study upon which a given model has not been trained will be referred to as ``unseen study'' data. 

\subsection{Predictive Models}
To predict fMRI QC labels, four different predictive models are evaluated using the \textit{sci-kit learn} Python library; Logistic Regression, Support Vector Machines (SVM), Random Forest, and Gradient Boosting classifiers. We tune the hyperparameters for the SVM, Random Forest, and Gradient Boosting models using 5-Fold Grid Search Cross Validation. 

\begin{table}[ht]
    \centering
    \resizebox{\linewidth}{!}{
        \def\arraystretch{1.25}
        \begin{tabular}{|c|c|c|c|c|c|c|}
        \hline
        \cellcolor{red!0!green!30!blue!30!} & \multicolumn{3}{|c|}{\cellcolor{red!0!green!30!blue!30!}\textbf{EMBARC}} & \multicolumn{3}{|c|}{\cellcolor{red!0!green!30!blue!30!}\textbf{CNP}} \\
        \hline
        \cellcolor{red!0!green!30!blue!30!}\textbf{Feature Set} & \cellcolor{red!0!green!30!blue!30!}\textbf{Classifier w/} & \cellcolor{red!0!green!30!blue!30!}\textbf{\# of Features} & \cellcolor{red!0!green!30!blue!30!}\textbf{Max} & \cellcolor{red!0!green!30!blue!30!}\textbf{Classifier w/} & \cellcolor{red!0!green!30!blue!30!}\textbf{\# of Features} & \cellcolor{red!0!green!30!blue!30!}\textbf{Max} \\
        \cellcolor{red!0!green!30!blue!30!}&
        \cellcolor{red!0!green!30!blue!30!}\textbf{Max AUC}&
        \cellcolor{red!0!green!30!blue!30!}\textbf{@ Max AUC} &\cellcolor{red!0!green!30!blue!30!}\textbf{AUC} & 
        \cellcolor{red!0!green!30!blue!30!}\textbf{Max AUC}&
        \cellcolor{red!0!green!30!blue!30!}\textbf{@ Max AUC} &\cellcolor{red!0!green!30!blue!30!}\textbf{AUC} \\
        \hline
        \textbf{FLAG-QC} & Random & 11 & 0.89 & SVM & 7 & 0.93 \\
         & Forest & & & & & \\
        \hline
        \textbf{MRIQC,} & Logistic & 18 & 0.86 & SVM & 13 & 0.79 \\
        \textbf{Functional} & Regression & & & & & \\
        \hline
        \textbf{MRIQC,} & Gradient & 26 & 0.86 & Random & 10 & 0.85 \\
        \textbf{Structural} & Boosting & &  & Forest & & \\
        \hline
        \textbf{All Features} & Random & 20 & 0.90 & SVM & 9 & 0.97 \\
        & Forest & & & & & \\
        \hline
        \end{tabular}
    }
    \caption{Summary of Within Dataset Forward Feature Selection Classification Results}
    \label{Tab:summary_sfs_results}
\end{table}

\begin{figure*}[hbt!]
    \centering
    ROC Curves \\
        \includegraphics[width=.48\linewidth]{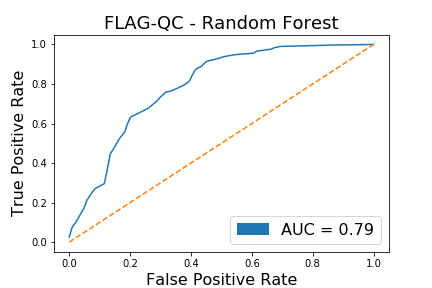}
        \centering
        \includegraphics[width=.48\linewidth]{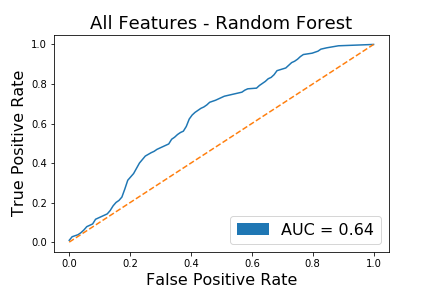}
    \\
    \hfill \\
        \centering
        \includegraphics[width=.48\linewidth]{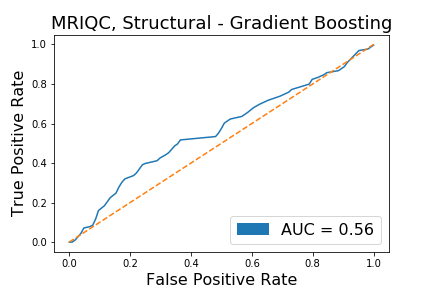}
        \centering
        \includegraphics[width=.48\linewidth]{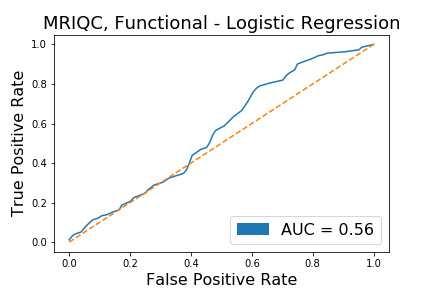} \\
        \smallskip
        \def\arraystretch{1.25}
        
        \begin{tabular}{|c|c|c|c|c|}
        \multicolumn{5}{c}{\Large{\textbf{Classification Metrics}}} \\
        \hline
        \cellcolor{red!0!green!30!blue!30!}\textbf{Metric} & \cellcolor{red!0!green!30!blue!30!}\textbf{FLAG-QC} & \cellcolor{red!0!green!30!blue!30!}\textbf{MRIQC} & \cellcolor{red!0!green!30!blue!30!}\textbf{MRIQC} & \cellcolor{red!0!green!30!blue!30!}\textbf{All Features} \\
        \cellcolor{red!0!green!30!blue!30!} & \cellcolor{red!0!green!30!blue!30!}\textbf{Logs} & \cellcolor{red!0!green!30!blue!30!}\textbf{Functional} & \cellcolor{red!0!green!30!blue!30!}\textbf{Structural} & \cellcolor{red!0!green!30!blue!30!} \\
        \hline
         AUC & 0.79 & 0.56 & 0.56 & 0.64 \\
        \hline
         Accuracy & 74.90\% & 61.35\% & 56.57\% & 64.54.\% \\
        \hline
         Precision & 0.72 & 0.62 & 0.59  & 0.64 \\
        \hline
         Recall & 0.95 & 0.85 & 0.83  & 0.93 \\
        \hline
        \end{tabular}
    \captionsetup{justification=centering}
    \caption{Prediction on Unseen Study data: Training Set - EMBARC, Test Set - CNP }
    \label{fig:final_roc_curves}
\end{figure*}

\subsection{Within Dataset Cross-Validation}

We first attempt to predict manual QC labels for held out sets of scans within datasets collected in a single study. We train and test Logistic Regression, SVM, Random Forest, and Gradient Boosting classifiers separately on each of the three feature sets mentioned before, labelled ``FLAG-QC'', ``MRIQC, Functional'', and ``MRIQC, Structural'', as well as the ensemble of all, labelled ``All Features''. To do so, we evaluate each feature-model pair within a 5-fold Cross Validation scheme, first using HSIC Lasso to reduce the dimensionality of the feature space. We then run Forward Feature Selection and report the mean AUC across folds for each number of selected features. The results using these methods for the EMBARC dataset are shown in Figure \ref{fig:emb_sfs_results}, and summary results for both EMBARC and CNP are reported in Table \ref{Tab:summary_sfs_results}.

In the EMBARC dataset, after Forward Feature Selection, we find that our FLAG-QC feature set achieves an AUC of 0.89. The other individual feature sets perform slightly worse with the AUC being 0.86 for the MRIQC, Functional features and 0.86 for the MRIQC, Structural features. We see that using all of the features together creates the classifier with the best performance, achieving an AUC of 0.90. Although, there is variability in which model performed best on each feature set, all models performed reasonably well across all feature sets, with the lowest feature-model AUC being 0.83 (MRIQC, Structural - SVM).

We replicate the same procedure on the CNP dataset, resulting in an AUC of 0.93 for the FLAG-QC (SVM); 0.79 for the MRIQC, Functional Features (SVM); 0.85 for MRIQC, Structural (Random Forest); and 0.97 for the ensemble feature sets (SVM). We see a similar pattern that the FLAG-QC features outperform the MRIQC feature sets (though by a larger magnitude this time), and the combination of all feature sets outperforms any individual set. Also again all feature-model pairs perform reasonably accurately (min AUC 0.77 with MRIQC, Functional features using Gradient Boosting). 

\subsection{Unseen Study Dataset as Test Set}

In our next experiment, we apply the same modeling framework to predict QC labels on one dataset, while the classifier is trained on data collected from a completely separate study. All 612 labelled scans from the EMBARC dataset are used as our training set. We use our results from EMBARC within dataset cross validation prediction with Forward Feature Selection to select the model that will be evaluated on the test set, CNP. For each feature set, we select the classifier with the highest AUC. The classifiers selected for each feature set are shown in Table \ref{Tab:summary_sfs_results}.

Within each feature set, we again start by running HSIC Lasso on the training dataset for an initial model independent feature selection, and then perform Forward Feature Selection to choose the final set of features to be tested on CNP data. Finally, we perform one last 5-fold CV parameter grid search to tune and train the model specifically for the final selected set of features. Using this framework, we predict manual QC labels on scans from the CNP dataset to evaluate our model's performance. 

The FLAG-QC features perform much better when predicting on the unseen study data from the CNP dataset than any other set of features, attaining an AUC of 0.79. The ROC curves from these predictions shown in Figure \ref{fig:final_roc_curves} clearly display a difference in performance between our novel feature set and those previously proposed. The individual MRIQC feature sets perform much worse on the unseen study, each only reaching an AUC of 0.56. Additionally, the second best performing set of features is ``All Features'' with an AUC of 0.64. This set of course contains the FLAG-QC features, further highlighting the importance of the FLAG-QC features in the classifiers' ability to generalize across datasets. The pronounced drop in performance in unseen study prediction associated with all models that include MRIQC features implies that these features may lead to greater overfitting on the training set as compared to that of FLAG-QC. The results achieved using the FLAG-QC features demonstrates our classifier's generalizability in predicting fMRI QC labels in unseen studies, a main goal of this work.


\section{Conclusion}

In our study we produced a number of novel findings. We demonstrate the first successful classifiers that predict manual QC labels on fMRI scans. Additionally, we develop a novel set of features that proves useful in making these QC predictions. And finally, our classifiers generalize to an unseen dataset with high accuracy and do so only when trained on our novel feature set. 

We'd like to note that the across-dataset generalizability of this method was only able to be tested on one unseen dataset. These methods should be further tested on additional datasets to confirm the robustness of this generalization.

In order to realize the full potential of automatic fMRI QC, we see a need to evaluate its impact. Before automatic QC is actively used in fMRI research we must assess how applying these techniques instead of manual QC affects the outcomes of later fMRI analysis. A clear question for further work arises: What is the performance bar for automatic QC at which subsequent analysis will yield similar results to data that has undergone manual QC? Answering questions like this will be vital to the practical implementation of automatic QC for fMRI analysis.

Automating the QC of fMRI data has the potential to meaningfully reduce the time spent by expert reviewers labelling the quality of these scans, unlocking new magnitudes of analysis. The impact of these methods are continually magnified by the growing scale of collected fMRI datasets. The weeks to months spent labelling datasets of this scale can instead be spent analyzing this powerful data modality that is steadily proving its ability to help researchers understand the brain basis of human health and disease. Our work demonstrates the feasibility of this paradigm, opening the door to exciting new insights generated from fMRI data toward the advancement of cognitive neuroscience, clinical psychiatry, and precision medicine.

\section{Acknowledgements}

We want to thank Humberto Gonzalez, Parvez Ahammad, and Yuelu Liu for their important discussions in the conceptualization of the automation of fMRI QC. Additionally, Yuelu Liu was instrumental in the design of the manual fMRI QC process discussed in this work.

\bibliographystyle{aaai}
\bibliography{Bibliography-File}

\begin{thebibliography}{}

\bibitem[\protect\citeauthoryear{Alfaro-Almagro \bgroup et al\mbox.\egroup
  }{2018}]{biobank}
Alfaro-Almagro, F.; et~al.
\newblock 2018.
\newblock Image processing and quality control for the first 10,000 brain
  imaging datasets from uk biobank.
\newblock {\em NeuroImage} 166:400 -- 424.

\bibitem[\protect\citeauthoryear{Casey \bgroup et al\mbox.\egroup
  }{2018}]{CASEY201843}
Casey, B.; et~al.
\newblock 2018.
\newblock The adolescent brain cognitive development (abcd) study: Imaging
  acquisition across 21 sites.
\newblock {\em Developmental Cognitive Neuroscience} 32:43 -- 54.
\newblock The Adolescent Brain Cognitive Development (ABCD) Consortium:
  Rationale, Aims, and Assessment Strategy.

\bibitem[\protect\citeauthoryear{Cox}{1996}]{COX1996162}
Cox, R.~W.
\newblock 1996.
\newblock Afni: Software for analysis and visualization of functional magnetic
  resonance neuroimages.
\newblock {\em Computers and Biomedical Research} 29(3):162 -- 173.

\bibitem[\protect\citeauthoryear{Cremers, Wager, and Yarkoni}{2017}]{cremers}
Cremers, H.~R.; Wager, T.~D.; and Yarkoni, T.
\newblock 2017.
\newblock The relation between statistical power and inference in fmri.
\newblock {\em PLOS ONE} 12(11):1--20.

\bibitem[\protect\citeauthoryear{Di~Martino \bgroup et al\mbox.\egroup
  }{2014}]{DiMartino2014}
Di~Martino, A.; et~al.
\newblock 2014.
\newblock The autism brain imaging data exchange: towards a large-scale
  evaluation of the intrinsic brain architecture in autism.
\newblock {\em Molecular Psychiatry} 19(6):659--667.

\bibitem[\protect\citeauthoryear{Drysdale \bgroup et al\mbox.\egroup
  }{2016}]{Drysdale2016}
Drysdale, A.~T.; et~al.
\newblock 2016.
\newblock Resting-state connectivity biomarkers define neurophysiological
  subtypes of depression.
\newblock {\em Nature Medicine} 23:28 EP --.
\newblock Article.

\bibitem[\protect\citeauthoryear{Essen \bgroup et al\mbox.\egroup
  }{2012}]{VANESSEN20122222}
Essen, D.~V.; et~al.
\newblock 2012.
\newblock The human connectome project: A data acquisition perspective.
\newblock {\em NeuroImage} 62(4):2222 -- 2231.
\newblock Connectivity.

\bibitem[\protect\citeauthoryear{Esteban \bgroup et al\mbox.\egroup
  }{2017}]{mriqc}
Esteban, O.; et~al.
\newblock 2017.
\newblock Mriqc: Advancing the automatic prediction of image quality in mri
  from unseen sites.
\newblock {\em PLOS ONE} 12(9):1--21.

\bibitem[\protect\citeauthoryear{Fischl \bgroup et al\mbox.\egroup
  }{2002}]{FISCHL2002341}
Fischl, B.; et~al.
\newblock 2002.
\newblock Whole brain segmentation: Automated labeling of neuroanatomical
  structures in the human brain.
\newblock {\em Neuron} 33(3):341 -- 355.

\bibitem[\protect\citeauthoryear{Gao, Calhoun, and
  Sui}{2018}]{ml_in_depression}
Gao, S.; Calhoun, V.~D.; and Sui, J.
\newblock 2018.
\newblock Machine learning in major depression: From classification to
  treatment outcome prediction.
\newblock {\em CNS Neuroscience \& Therapeutics} 24(11):1037--1052.

\bibitem[\protect\citeauthoryear{Hastie, Tibshirani, and Friedman}{2009}]{sfs}
Hastie, T.; Tibshirani, R.; and Friedman, J.
\newblock 2009.
\newblock {\em The Elements of Statistical Learning: Data Mining, Inference,
  and Prediction.}
\newblock Springer.

\bibitem[\protect\citeauthoryear{Liu \bgroup et al\mbox.\egroup
  }{2018}]{Liu453951}
Liu, Y.; et~al.
\newblock 2018.
\newblock Highly predictive transdiagnostic features shared across
  schizophrenia, bipolar disorder, and adhd identified using a machine learning
  based approach.
\newblock {\em bioRxiv}.

\bibitem[\protect\citeauthoryear{Liu \bgroup et al\mbox.\egroup
  }{2019}]{liu_ml_in_mdd}
Liu, Y.; et~al.
\newblock 2019.
\newblock Machine learning identifies large-scale reward-related activity
  modulated by dopaminergic enhancement in major depression.
\newblock {\em Biological Psychiatry: Cognitive Neuroscience and Neuroimaging}.

\bibitem[\protect\citeauthoryear{Lu \bgroup et al\mbox.\egroup
  }{2019}]{QIMS25794}
Lu, W.; Dong, K.; Cui, D.; Jiao, Q.; and Qiu, J.
\newblock 2019.
\newblock Quality assurance of human functional magnetic resonance imaging: a
  literature review.
\newblock {\em Quantitative Imaging in Medicine and Surgery} 9(6).

\bibitem[\protect\citeauthoryear{Mellem \bgroup et al\mbox.\egroup
  }{2018}]{Mellem414037}
Mellem, M.~S.; et~al.
\newblock 2018.
\newblock Machine learning models identify multimodal measurements highly
  predictive of transdiagnostic symptom severity for mood, anhedonia, and
  anxiety.
\newblock {\em bioRxiv}.

\bibitem[\protect\citeauthoryear{Mortamet \bgroup et al\mbox.\egroup
  }{2009}]{mortamet}
Mortamet, B.; et~al.
\newblock 2009.
\newblock Automatic quality assessment in structural brain magnetic resonance
  imaging.
\newblock {\em Magnetic Resonance in Medicine} 62(2):365--372.

\bibitem[\protect\citeauthoryear{Pizarro \bgroup et al\mbox.\egroup
  }{2016}]{pizzaro}
Pizarro, R.~A.; et~al.
\newblock 2016.
\newblock Automated quality assessment of structural magnetic resonance brain
  images based on a supervised machine learning algorithm.
\newblock {\em Frontiers in Neuroinformatics} 10:52.

\bibitem[\protect\citeauthoryear{Poldrack \bgroup et al\mbox.\egroup
  }{2016}]{CNP}
Poldrack, R.~A.; et~al.
\newblock 2016.
\newblock A phenome-wide examination of neural and cognitive function.
\newblock {\em Scientific Data} 3(1):160110.

\bibitem[\protect\citeauthoryear{Reuter}{2013}]{reuter_2013}
Reuter, M.
\newblock 2013.
\newblock Freesurfer.

\bibitem[\protect\citeauthoryear{Soares \bgroup et al\mbox.\egroup
  }{2016}]{hitchhiker_fmri}
Soares, J.~M.; et~al.
\newblock 2016.
\newblock A hitchhiker's guide to functional magnetic resonance imaging.
\newblock {\em Frontiers in Neuroscience} 10:515.

\bibitem[\protect\citeauthoryear{Trivedi \bgroup et al\mbox.\egroup
  }{2016}]{embarc}
Trivedi, M.~H.; et~al.
\newblock 2016.
\newblock Establishing moderators and biosignatures of antidepressant response
  in clinical care (embarc): Rationale and design.
\newblock {\em Journal of Psychiatric Research} 78:11 -- 23.

\bibitem[\protect\citeauthoryear{Woodard and
  Carley-Spencer}{2006}]{Woodard2006}
Woodard, J.~P., and Carley-Spencer, M.~P.
\newblock 2006.
\newblock No-reference image quality metrics for structural mri.
\newblock {\em Neuroinformatics} 4(3):243--262.

\bibitem[\protect\citeauthoryear{Yamada \bgroup et al\mbox.\egroup
  }{2014}]{Yamada_2014}
Yamada, M.; Jitkrittum, W.; Sigal, L.; Xing, E.~P.; and Sugiyama, M.
\newblock 2014.
\newblock High-dimensional feature selection by feature-wise kernelized lasso.
\newblock {\em Neural Computation} 26(1):185–207.

\end{thebibliography}

\end{document}